\documentclass[]{spie}

\usepackage{amsmath,amsfonts,amssymb}
\usepackage{graphicx}
\usepackage[colorlinks=true, allcolors=blue]{hyperref}
\usepackage{siunitx}
\usepackage{multirow}
\usepackage[table,xcdraw]{xcolor}
\usepackage{lineno}

\usepackage{soul}

\title{Characterization of a 15 $\mu m$ Cutoff HgCdTe Detector Array for Astronomy}

 \author[a,*]{Mario S. Cabrera}
 \author[a]{Craig W. McMurtry}
 \author[a]{William J. Forrest}
 \author[a]{Judith L. Pipher}
 \author[a]{Meghan L. Dorn}
 \author[b]{Donald Lee}
 \affil[a]{University of Rochester, Department of Physics and Astronomy, 500 Wilson Blvd., Rochester, NY, USA, 14627-0171}
 \affil[b]{Teledyne Imaging Sensors, 5212 Verdugo Way, Camarillo, CA, USA, 93012}

\authorinfo{*Mario S. Cabrera: mcabrer2@ur.rochester.edu}

\begin{document}

\maketitle

\begin{abstract}
The University of Rochester infrared detector group is working together with Teledyne Imaging Sensors to develop HgCdTe 15 $\mu m$ cutoff wavelength detector arrays for future space missions. To reach the 15 $\mu m$ cutoff goal, we took an intermediate step by developing four $\sim$13 $\mu m$ cutoff wavelength arrays to identify any unforeseen effects related to increasing the cutoff wavelength from the extensively characterized 10 $\mu m$ cutoff wavelength detector arrays developed for the NEOCam mission. The characterization of the $\sim$13 $\mu m$ cutoff wavelength HgCdTe arrays at the University of Rochester allowed us to determine the key dark current mechanisms that limit the performance of these HgCdTe detector arrays at different temperatures and bias when the cutoff wavelength is increased. We present initial dark current and well depth measurements of a 15 $\mu m$ cutoff array which shows dark current values two orders of magnitude smaller at large reverse bias than would be expected from our previous best structures.
\end{abstract}


\section{Introduction}
\label{sect:intro}

The University of Rochester infrared detector group is working together with Teledyne Imaging Sensors (TIS) to develop HgCdTe 15 $\mu m$ cutoff wavelength detector arrays for future space missions. The target cutoff wavelength goal of 15 $\mu m$ for this project was chosen to demonstrate the performance of HgCdTe detector arrays as a viable option for future missions aimed at studying the atmospheres of exoplanets. Future missions developed to study exoplanets would benefit from these detector arrays because solar systems with planets (or forming planets) can be detected with far better contrast at infrared rather than visible wavelengths. This technology would also enable the detection of $CO_2$ at 15 $\mu m$, a signature indicative of a terrestrial planet in the habitable zone\cite{Kaltenegger17}.

Efforts in the development and improvement of HgCdTe detector arrays with cutoff wavelengths longer than $\sim$ 5.4 $\mu m$ has been carried out by the University of Rochester infrared detector team since 1992.\cite{Wu97,Bailey98,Bacon04,Bacon06,Bacon10} The first deliveries of HgCdTe single-photodiodes\cite{Wu97} to UR by Rockwell with cutoff wavelength of 10.6 $\mu m$ (at a temperature of 40 K) had dark currents $>10^3\ e^-/sec$ at temperatures between 20 and 40 K with a reverse bias of 20 mV.

By 2003, early large format arrays had 512$\times$512 pixels. Two of these first large format arrays with cutoff wavelengths of 9.3 and 10.3 $\mu m$\cite{Bacon10}, showed low dark current, but could not support a large reverse bias, which limited the well depth of these devices. Three other arrays with cutoff wavelengths ranging from 8.4 to 9.1 $\mu m$, showed dark currents below 30 $e^-/sec$ and well depths of at least 40 mV at a temperature of $\sim$30 K for more than 70\% of pixels\cite{Bacon06}. Even with these limitations, the devices showed promise as they were operable at a temperature of 30 K. The latest $\sim$10 $\mu m$ HgCdTe cutoff wavelength devices developed for the proposed NEOCam mission have 2048$\times$2048 pixel format, and 98.9\% of pixels have shown to have dark currents $<$200 $e^-/s$ and well depths $>$44,000 $e^-$ with an applied reverse bias of 250 mV at a temperature of 40 K\cite{Dorn18} (median dark current and well depth of 0.3 $e^-/s$ and 65,000 $e^-$ respectively). Increasing the cutoff wavelength of these devices has proven to not be a trivial process, and the lessons learned from each iteration of the LW10 devices have led to the improvement of these devices.

The first step of this project to extend the cutoff wavelength to 15 $\mu m$ (LW15 arrays) was to first develop HgCdTe detector arrays with a cutoff wavelength of $\sim$ 13 $\mu m$ (LW13 arrays) with different array designs to mitigate dark currents, namely the expected increase in tunneling currents due to the decrease in bandgap energy relative to the LW10 devices. This step was crucial to identify the best array design from TIS that would best guarantee the success of further increasing the cutoff wavelength to 15 $\mu m$.

\subsection{LW13 Phase Summary}
\label{sect:LW13 summary}

TIS delivered four 1024$\times$1024 pixel detector arrays to the University of Rochester Infrared Detector group for characterization. Hawaii-1RG (H1RG) multiplexers\cite{Montroy02,Loose02,Loose07} were chosen for the entire project because of the low power dissipation, ideal for passively cooled space missions, in addition to the small spread in the zero-bias point compared to other muxes. All four detector arrays had cutoff wavelengths ranging from 12.4 to 12.8 $\mu m$, measured at a temperature of 30 K. Two of the arrays (H1RG-18367 and 18508) had the same design as the LW10 detector arrays designed for the proposed NEOCam mission\cite{Dorn16,McMurtry16,Dorn18} extrapolated to longer wavelengths by decreasing the mole fraction of cadmium to mercury in the HgCdTe alloy\cite{Hansen83}, while the other two arrays (H1RG-18369 and 18509) had two different TIS proprietary designs with the goal of mitigating quantum tunneling dark currents\cite{Kinch14,Kinch81,Sze69,Bacon06}.

The median dark current per pixel for three of the four arrays was below 1 $e^-/s$ at a temperature of 28 K and 150 mV of applied reverse bias, with a median well depth of $\sim$ 43 $ke^-$. We were able to show that the dark current is dominated by G-R\cite{Sah57} and diffusion\cite{Reine81} as the operating temperature is increased, while increasing the bias (for larger well depth) increases quantum tunneling dark currents exponentially. The theory of these dark current mechanisms will be discussed in section \ref{sect:theory}.

LW13 array H1RG-18509, designed to mitigate the effects of quantum tunneling dark currents, had the best dark current and well depth performance at larger applied reverse biases. At a temperature of 28 K and applied bias of 350 mV, 86\% of the pixels had dark currents below 10 $e^-/s$ and well depth of at least 75 $ke^-$ (median dark current and well depth of 1.8$e^-/s$ and 81 $ke^-$ respectively). The other three LW13 arrays H1RG-18367, 18369, and 18508 had median dark currents of about 379, 730, and 780 $e^-/s$ at 28 K and applied bias of 350 mV, where the dark current at this bias is dominated by band-to-band or trap-to-band (defect assisted) tunneling\cite{Cabrera19}.

An increase in band-to-band and defect assisted tunneling current was expected array-wide as the cutoff wavelength is increased to a target of 15 $\mu m$ due to the smaller band-gap. The effect of trap-to-band is less predictable than band-to-band tunneling since it is dependent on the defect/dislocation density of individual pixels (see section \ref{sect:theory}). 

The characterization results from the LW13 arrays suggests that a similar array design used for H1RG-18509 would be the best approach for the final phase of the project to extend the cutoff wavelength to 15 $\mu m$ to reduce quantum tunneling dark currents.

\subsection{LW15 Arrays}
\label{sect:LW15 arrays}

Our group has received three detector arrays (H1RG-20302, H1RG-20303, and H1RG-20304) for the final phase of the project from TIS. Table \ref{tab: QE and cutoff} includes the quantum efficiency (QE) and cutoff wavelength measurements provided by TIS for the three LW15 arrays at a temperature of 30 K from Process Evaluation Chips (PECs). The PECs were grown and processed at the same time as the megapixel arrays. We have tested so far one of the three arrays, H1RG-20303, and the preliminary calibration and characterization of the dark current are presented here.

\begin{table}[htb!]
\centering
\caption{Cutoff wavelength and QE measurements for all three LW15 arrays were provided by TIS at a temperature of 30 K with an applied reverse bias of 100 mV. These arrays do not have anti-reflective coating. The PEC QE measurements without anti-reflective coating are above the theoretical value (78\%), but are within experimental measurement error.}
\label{tab: QE and cutoff}
\begin{tabular}{|c|c|c|c|}
\hline
\begin{tabular}[c]{@{}c@{}}Detector\\ H1RG-\end{tabular} & Wafer & \begin{tabular}[c]{@{}c@{}}Cutoff\\ Wavelength\\ ($\mu m$)\end{tabular} & \begin{tabular}[c]{@{}c@{}}QE\\ (6-12 $\mu m$)\end{tabular} \\ \hline
20302 & 3995 & 16.7 & 81\% \\ \hline
20303 & 3994 & 15.5 & 83\% \\ \hline
20304 & 4018 & 15.2 & 80\% \\ \hline
\end{tabular}
\end{table}

\section{Data Acquisition}
\label{data}

To characterize the performance of H1RG-20303, dark current and well depth measurements were obtained per pixel at stable temperatures ranging from 23 to 30 K with applied reverse bias of 50, 150, 250, and 350 mV. In addition to the stable temperature dark current, we measured the dark current with slowly increasing temperature starting at 25 K up to a temperature of 37 K with an applied bias of 150 mV to fit dark current models to our data. The lower limit on the operating temperature of 23 K was to minimize the dark current due to thermal processes as much as possible. At 23 K, it will be shown in section \ref{sect:results} that at this temperature, tunneling currents have a significant contribution to the total dark current, and will not continue to decrease with decreasing temperature. Additionally, the H1RG multiplexer is designed to operate between 30-300 K\cite{Loose07}, where the read out noise in the multiplexer can increase with decreasing temperature due to freeze outs.\cite{Fossum93} 

\subsection{Dark Current and Well Depth}
\label{dark current data acq}

The dark current is measured by taking 200 non-destructive samples-up-the-ramp (SUTR)\cite{Fowler90,Garnett93,Rauscher07} in the dark with an integration time of 5.8 seconds between samples, where the dark current per pixel that we present here corresponds to the slope at the beginning of the signal \textit{vs.} time curve (see Fig. \ref{fig:sutr}).

Following the reset of the array, a redistribution of charge due to the capacitive coupling to the reset FET can add 0-75 mV to the applied reverse bias\cite{Bacon06,McMurtry13,Cabrera19}. The initial dark current can change considerably with respect to the detector bias on each individual pixel at the beginning of the integration ramp: we therefore measure the zero bias saturation level on the same ramp used to measure the dark current without resetting the array between measurements, allowing us to obtain the actual detector bias. The well depth is given by the detector bias for the first sample.

\subsection{Dark Current \textit{vs.} Bias}
\label{I-V}

The data set taken to determine the initial dark current and well depth also allows us to determine the dark current and detector bias per pixel at any time in the dark integration ramp as pixels debias. The dark current \textit{vs.} bias (I-V) curve is given by the slope of the SUTR data versus the detector bias voltage (e.g. Fig. \ref{fig:sutr}).

\subsection{Dark Current \textit{vs.} Temperature}
\label{warmup}

The I-V and dark current \textit{vs.} temperature (I-T) curves are used to compare the dark current behavior observed in this array with theory of several dark current mechanisms such as diffusion, G-R, band-to-band and trap-to-band tunneling.

The I-T curve is composed of the initial dark current measured at stable temperatures from 23-30 K and dark current measured at varying temperatures when the liquid helium cryogen is exhausted. Dark exposures were continuously taken to measure the dark current as the temperature increased.

\section{Dark Current Theory}
\label{sect:theory}

The four main mechanisms of dark current that have been identified in the LW10, LW13, and now LW15 devices are diffusion, G-R, band-to-band quantum tunneling, and trap-to-band quantum tunneling dark currents. Diffusion and G-R are thermally generated dark currents, where diffusion dark currents are generated by electrons in the valence band gaining enough thermal energy to transition to the conduction band in the bulk of the material, while G-R dark currents occur in the depletion region as electrons transition from the valence to the conduction band via traps with energies that lie within the bandgap. 

Under the assumption that the diffusion length of minority carriers in the bulk of the material is larger than the thickness of this array, diffusion current is given by\cite{Reine81}
\begin{equation}
    \label{eq:diffusion}
    I_{dif} = A\frac{n_i^2 d}{N_d \tau_h }\left[exp\left(\frac{qV_{detector\ bias}}{k_b T}\right) - 1\right] ,
\end{equation}
where $A$ is the diode junction area, $d$ is the thickness of the n-type region, $N_d$ is the doping density, $\tau_h$ is the minority carrier (hole in the n-type region) lifetime, $k_b$ is Boltzmann's constant, T is the temperature, $V_{detector\ bias}$ is the detector reverse bias across the diode, and $n_i$ is the intrinsic carrier concentration given by\cite{Hansen83}
\begin{equation}
\label{eq:n_i}
n_i = \left( 5.585 - 3.820x + 1.753 \times 10^{-3}T - 1.364 \times 10^{-3}xT \right) \times \left[ 10^{14} E_g^{3/4} T^{3/2} exp\left( - \frac{E_g}{2k_b T} \right) \right],
\end{equation}
where $x$ is the cadmium mole fraction, and $E_g$ is the band gap energy (in eV) for HgCdTe given by\cite{Hansen83}
\begin{equation}
\label{eq:E_g}
E_g(x,T) = -0.302 + 1.93x - 0.81x^2 + 0.832x^3 + 5.35\times10^{-4}T(1 - 2x).
\end{equation}
The dependence on the square of the intrinsic carrier concentration makes diffusion dark currents a strong function of temperature.

G-R current is derived in Sah (1957)\cite{Sah57} and given by
\begin{equation}
\label{eq:g-r}
I_{G-R} = \frac{n_iW_DA}{\tau_{GR}}\left[\frac{sinh\left(\frac{-qV_{detector\ bias}}{2kT}\right)}{\frac{q\left(V_{bi}-V_{detector\ bias}\right)}{2kT}}\right]f(b),
\end{equation}

\begin{equation}
\label{eq:f}
f(b)=\int_{0}^{\infty}\frac{dz}{z^2+2bz+1},
\end{equation}

\begin{equation}
\label{eq:b}
b=exp\left[\frac{-qV_{detector\ bias}}{2kT}\right]\cosh\left(\frac{E_i-E_{t_{gr}}}{kT}\right).
\end{equation}
$E_{t_{gr}}$ is the trap energy level position with respect to the valence band which contributes the most to the G-R current, $E_i$ is the intrinsic Fermi level ($E_g/2$), $\tau_{GR}$ is the depletion region lifetime for holes and electrons, and $W_D$ is the depletion region width\cite{Wu97}. The weaker temperature dependence of G-R relative to diffusion means that at lower temperatures this dark current component can be considerably larger than diffusion. Both of these thermally generated dark currents are a weak function of bias in the regime where the arrays are characterized in our lab ($>$ 25 mV of reverse bias).

Tunneling currents are generated by the tunneling of electrons in the depletion region from the valence band to empty states in the conduction band directly (band-to-band) or indirectly via trap states in the bandgap (trap-to-band). The distance $\Delta z$ that the electrons have to tunnel through for a triangular barrier is\cite{Sze69} $\Delta z = E_g/qE$, where $E_g$ is the band gap energy, $q$ is the charge of an electron, and $E$ is the electric field $\left(\propto \sqrt{E_g - qV_{detector\ bias}}\right)$. Both tunneling currents are functions of the cutoff wavelength and bias, where increasing the cutoff wavelength (decreasing bandgap energy) and/or increasing the reverse bias will increase the tunneling probability by decreasing the distance that electrons need to tunnel. For a triangular barrier, both tunneling dark currents are given by\cite{Sze69, Kinch81, Kinch14}
\begin{equation}
\label{eq:band-to-band}
I_{band-to-band} = -\frac{q^2AEV_{detector\ bias}}{4\pi^2 \hbar^2} \sqrt{\frac{2m_{eff}}{E_g}} exp\left( -\frac{4\sqrt{2m_{eff}}E_g^{3/2}}{3q\hbar E} \right),
\end{equation}
\begin{equation}
\label{eq:trap-to-band}
I_{trap-to-band} = \frac{\pi^2 q m_{eff} A V_{detector\ bias} M^2 n_t}{h^3 \left( E_g - E_t\right)} exp \left( -\frac{4\sqrt{2 m_{eff}} \left(E_g-E_t\right)^{3/2}}{3q\hbar E} \right).
\end{equation}
$m_{eff}$ is the effective mass of the minority carrier, $M$ is the mass matrix, $E_t$ is the energy of the trap level with respect to the valence band, and $n_t$ is the trap density in the depletion region at $E_t$, and $E$ is the electric field across the depletion region given by\cite{Reine81}
\begin{equation}
\label{eq:E-field}
E = \sqrt{\frac{2N_d\left( E_g - qV_{detector\ bias} \right)}{\epsilon \epsilon_0}}.
\end{equation}

Data from the LW10\cite{Wu97,Bacon06,Bacon10} and the LW13\cite{Cabrera19} devices have shown that a single trap density is not sufficient to model trap-to-band tunneling. Pixels that exhibit trap-to-band tunneling in those devices may show a soft breakdown in the I-V curves, followed by the onset of trap-to-band tunneling. Similar behavior has been observed by other authors in 4H-SiC\cite{Neudeck98,Neudeck99} and Silicon\cite{Ravi73} diodes caused by screw dislocations and stacking faults respectively. The soft breakdown is caused by traps that become electrically active at a certain ``threshold voltage". The soft breakdowns shown in Neudeck et al.\cite{Neudeck98} and Ravi et al.\cite{Ravi73} were not uniform, and showed no correlation  between the onset of the soft breakdown and the trap density since the traps associated with dislocations may have different threshold voltages at which they become electrically active.

Bacon (2006)\cite{Bacon06} introduced a variable trap density to parametrically fit the soft breakdown and is given by
\begin{equation}
\label{eq:n_t}
n_t = n_{t_i} + \frac{n_{t_d}}{1 + exp \left[\frac{\gamma q\left(V_a + V_{detector\ bias}\right)}{kT} \right]},
\end{equation}
where $n_{t_i}$ is an initial active trap density, $n_{t_d}$ is the trap density due to activated dislocations at a voltage $V_a$, and $\gamma$ determines the rate at which the soft breakdown causes the current to increase before reaching the current expected from trap-to-band tunneling.

\section{Calibration}
\label{sect:calibration}

The first step in calibrating the data is to measure the source-follower FET gain in the multiplexer to convert between the output and input-referred signal. To do so, the reset switch is kept closed to vary the reset (input) voltage, while measuring the output voltage. The resulting gain had a value of 0.89.

Next, the nodal capacitance was measured using the $\sigma^2$ vs. signal method\cite{Mortara81}. This method takes advantage of the Poisson statistics nature of photons, where the noise in the incident photons (shot noise) for a given pixel is $\sqrt{S}$, where $S$ is the number of collected charges by a given pixel. If the array is operated in a regime in which the dominant component in the noise is the shot noise (i.e. photodiode signal is photon flux limited), using Poisson statistics, the total noise squared of the signal is
\begin{equation}
    \label{eq:Shot_noise_sqrd}
    \sigma ^2_{S} = S.
\end{equation}
The output signal is not the number of charges collected in the p-type implant, but the voltage at the integrating node ($V_{out}$) due to the collected charges. To convert the signal and noise due to the collected charges and the output voltage across a single pixel, the following relations are used:
\begin{gather}
    \label{eq:cap_volt_charge}
    V_{out}=\frac{Sq}{C} \\
    \sigma _{V_{out}}=\frac{\sigma _Sq}{C},
\end{gather}
where q is the elementary charge, and C is the pixel capacitance. Combining the above relations with equation \ref{eq:Shot_noise_sqrd}, the pixel capacitance is given by
\begin{equation}
    \label{eq:Capacitance}
    C=\frac{qV_{out}}{\sigma ^2_{V_{out}}}.
\end{equation}

\begin{figure}
\begin{center}
\begin{tabular}{c}
\includegraphics[scale=0.45]{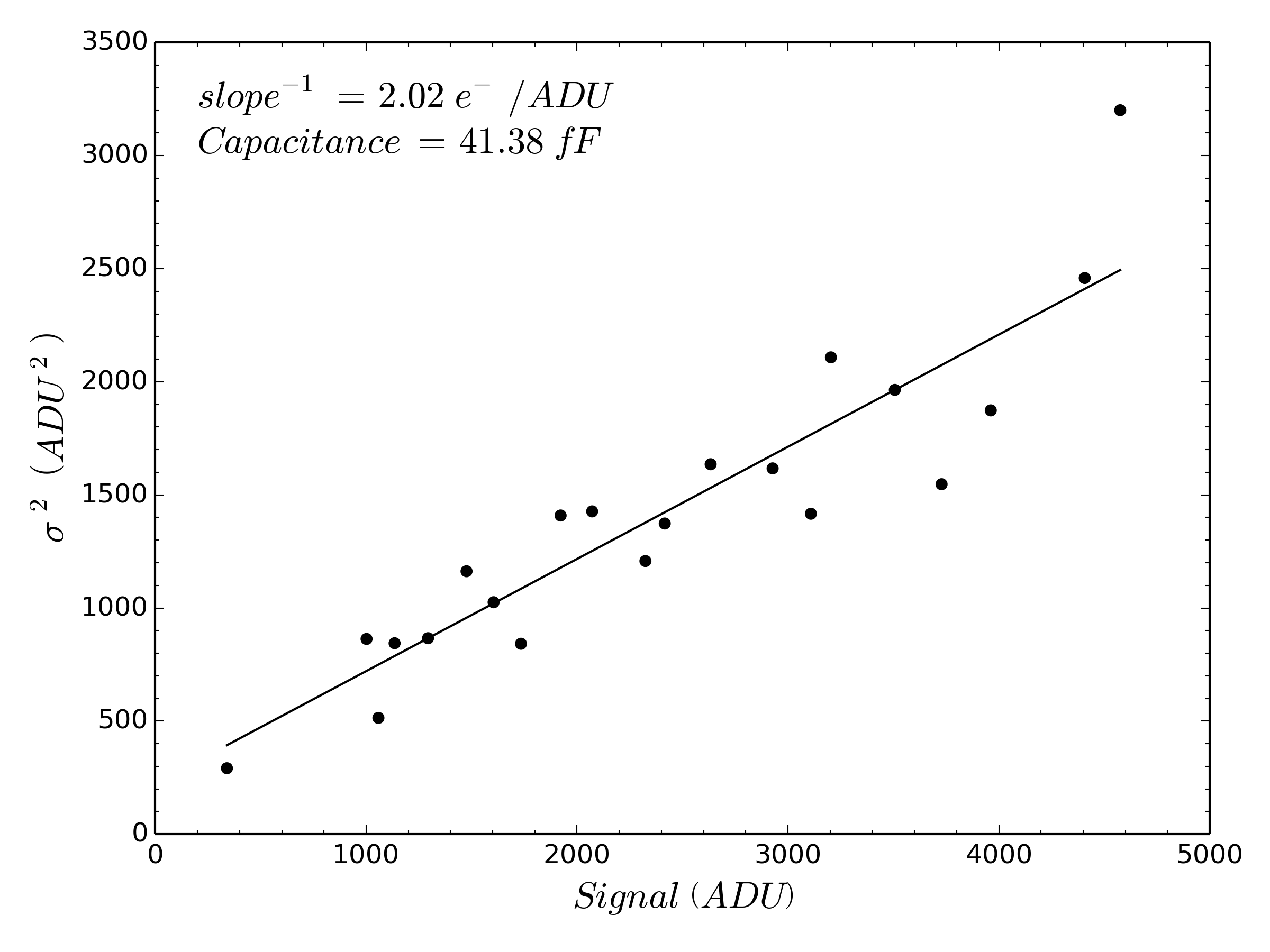}
\end{tabular}
\end{center}
\caption 
{ \label{fig:20303_sig2vssig}
Noise squared \textit{vs.} signal plot for a pixel in H1RG-20303 at a temperature of 25K and 50 mV of applied bias, where the slope of the fitted line corresponds to the conversion factor between ADUs and $e^-$. The capacitance calculated here is not yet corrected for interpixel capacitance.} 
\end{figure} 

\begin{figure}
\begin{center}
\begin{tabular}{c}
\includegraphics[scale=0.45]{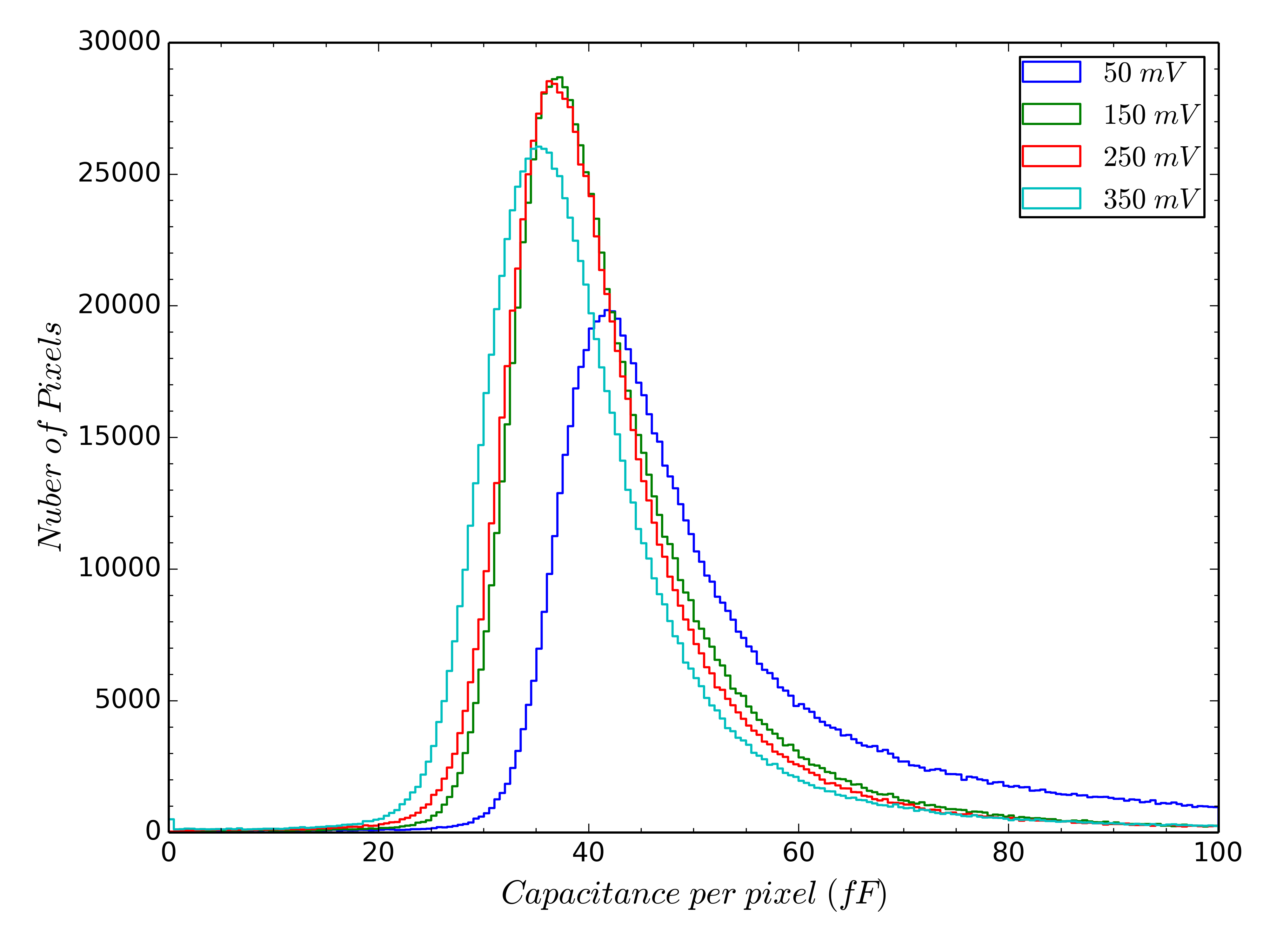}
\end{tabular}
\end{center}
\caption 
{ \label{fig:20303_capperpix}
Capacitance per pixel distribution for H1RG-20303 at a temperature of 25 K and at all tested applied biases. These capacitances have not yet been corrected for interpixel capacitance.} 
\end{figure}

Figure \ref{fig:20303_sig2vssig} shows the noise squared \textit{vs.} signal in analog to digital units (ADU), which can be converted to volts by dividing the 5 V range of the 16 bit A/D converter by $2^{16}$ ADUs and the gain from our array controller electronics. The inverse of the slope is then used to calculate the capacitance per pixel. Figure \ref{fig:20303_capperpix} shows the distribution of nodal capacitance per pixel at each of the four applied biases. The apparent spread in capacitance is mostly due to uncertainties in this method. The pixel capacitance using this method is overestimated, namely due to the interpixel capacitance (IPC) effect, which can capacitively couple signal between nearest neighboring pixels. Moore et al. (2004) showed this process will appear to add signal to a pixel, without adding noise.\cite{Moore04} The overestimation of the capacitance was shown explicitly in Finger et al. (2005)\cite{Finger2005} by measuring the nodal capacitance of CMOS arrays using the signal \textit{vs.} noise squared method and by a different direct measurement technique, where correcting for IPC can reduce the overestimation of the capacitance.

The capacitance values shown in Figures \ref{fig:20303_sig2vssig} and \ref{fig:20303_capperpix} are not yet corrected for IPC. The nearest neighbor method\cite{Moore04} was used to determine the IPC, where pixels with very high dark current were used to determine the coupling parameter $\alpha$. The nodal capacitance is then corrected by a factor of $1-8\alpha$. The alpha parameter determined for this array is 0.9\%, where the IPC corrected median nodal capacitance is 44, 37, 36, and 35 fF for applied biases of 50, 150, 250, and 350 mV respectively. The corrected nodal capacitance is used to determine the conversion between measured voltages and electrons.

\section{Results}
\label{sect:results}

H1RG-20303 was designed to reduce quantum tunneling dark currents, has a cutoff wavelength of 15.5 $\mu m$, and QE of 83\% between $\sim$ 6 to 12 $\mu m$ before anti-reflective coating.

\begin{table}
\centering
\caption{Median dark current and well depth with different applied biases and temperatures for H1RG-20303.}
\label{tab:median dark current and well depth}
\begin{tabular}{c|c|c|c|c|}
\cline{2-5}
\multicolumn{1}{l|}{} & Bias = 50 mV & 150 mV & 250 mV & 350 mV \\ \hline
\rowcolor[HTML]{C0C0C0} 
\multicolumn{1}{|c|}{\cellcolor[HTML]{C0C0C0}Temperature} & \multicolumn{4}{c|}{\cellcolor[HTML]{C0C0C0}\begin{tabular}[c]{@{}c@{}}Median Dark Current ($e^-/sec$)\\ Median Well Depth ($ke^-$, $mV$)\end{tabular}} \\ \hline
\rowcolor[HTML]{FFFFFF} 
\multicolumn{1}{|c|}{\cellcolor[HTML]{FFFFFF}} & 7 & 16 & 263 & 790 \\ \cline{2-5} 
\rowcolor[HTML]{FFFFFF} 
\multicolumn{1}{|c|}{\multirow{-2}{*}{\cellcolor[HTML]{FFFFFF}23 K}} & 20, 75 & 41, 176 & 62, 274 & 66, 306 \\ \hline
\rowcolor[HTML]{C0C0C0} 
\multicolumn{1}{|c|}{\cellcolor[HTML]{C0C0C0}} & 13 & 29 & 272 & 804 \\ \cline{2-5} 
\rowcolor[HTML]{C0C0C0} 
\multicolumn{1}{|c|}{\multirow{-2}{*}{\cellcolor[HTML]{C0C0C0}24 K}} & 20, 74 & 40, 175 & 62, 273 & 67, 308 \\ \hline
\rowcolor[HTML]{FFFFFF} 
\multicolumn{1}{|c|}{\cellcolor[HTML]{FFFFFF}} & 29 & 53 & 301 & 858 \\ \cline{2-5} 
\rowcolor[HTML]{FFFFFF} 
\multicolumn{1}{|c|}{\multirow{-2}{*}{\cellcolor[HTML]{FFFFFF}25 K}} & 20, 73 & 40, 174 & 62, 272 & 67, 311 \\ \hline
\rowcolor[HTML]{C0C0C0} 
\multicolumn{1}{|c|}{\cellcolor[HTML]{C0C0C0}} & 69 & 104 & 373 & 943 \\ \cline{2-5} 
\rowcolor[HTML]{C0C0C0} 
\multicolumn{1}{|c|}{\multirow{-2}{*}{\cellcolor[HTML]{C0C0C0}26 K}} & 20, 72 & 40, 173 & 61, 270 & 68, 314 \\ \hline
\rowcolor[HTML]{FFFFFF} 
\multicolumn{1}{|c|}{\cellcolor[HTML]{FFFFFF}} & 179 & 205 & 478 & 1080 \\ \cline{2-5} 
\rowcolor[HTML]{FFFFFF} 
\multicolumn{1}{|c|}{\multirow{-2}{*}{\cellcolor[HTML]{FFFFFF}27 K}} & 19, 69 & 39, 169 & 61, 267 & 68, 313 \\ \hline
\rowcolor[HTML]{C0C0C0} 
\multicolumn{1}{|c|}{\cellcolor[HTML]{C0C0C0}} & 350 & 427 & 649 & 1307 \\ \cline{2-5} 
\rowcolor[HTML]{C0C0C0} 
\multicolumn{1}{|c|}{\multirow{-2}{*}{\cellcolor[HTML]{C0C0C0}28 K}} & 18, 65 & 38, 165 & 59, 261 & 67, 310 \\ \hline
\rowcolor[HTML]{FFFFFF} 
\multicolumn{1}{|c|}{\cellcolor[HTML]{FFFFFF}} & 646 & 837 & 1107 & 1712 \\ \cline{2-5} 
\rowcolor[HTML]{FFFFFF} 
\multicolumn{1}{|c|}{\multirow{-2}{*}{\cellcolor[HTML]{FFFFFF}29 K}} & 16, 60 & 37, 159 & 57, 253 & 66, 305 \\ \hline
\rowcolor[HTML]{C0C0C0} 
\multicolumn{1}{|c|}{\cellcolor[HTML]{C0C0C0}} & 1275 & 1848 & 2135 & 2606 \\ \cline{2-5} 
\rowcolor[HTML]{C0C0C0} 
\multicolumn{1}{|c|}{\multirow{-2}{*}{\cellcolor[HTML]{C0C0C0}30 K}} & 14, 53 & 35, 150 & 55, 240 & 63, 292 \\ \hline
\end{tabular}
\end{table}

\begin{figure}
\begin{center}
\begin{tabular}{c}
\includegraphics[scale=0.4]{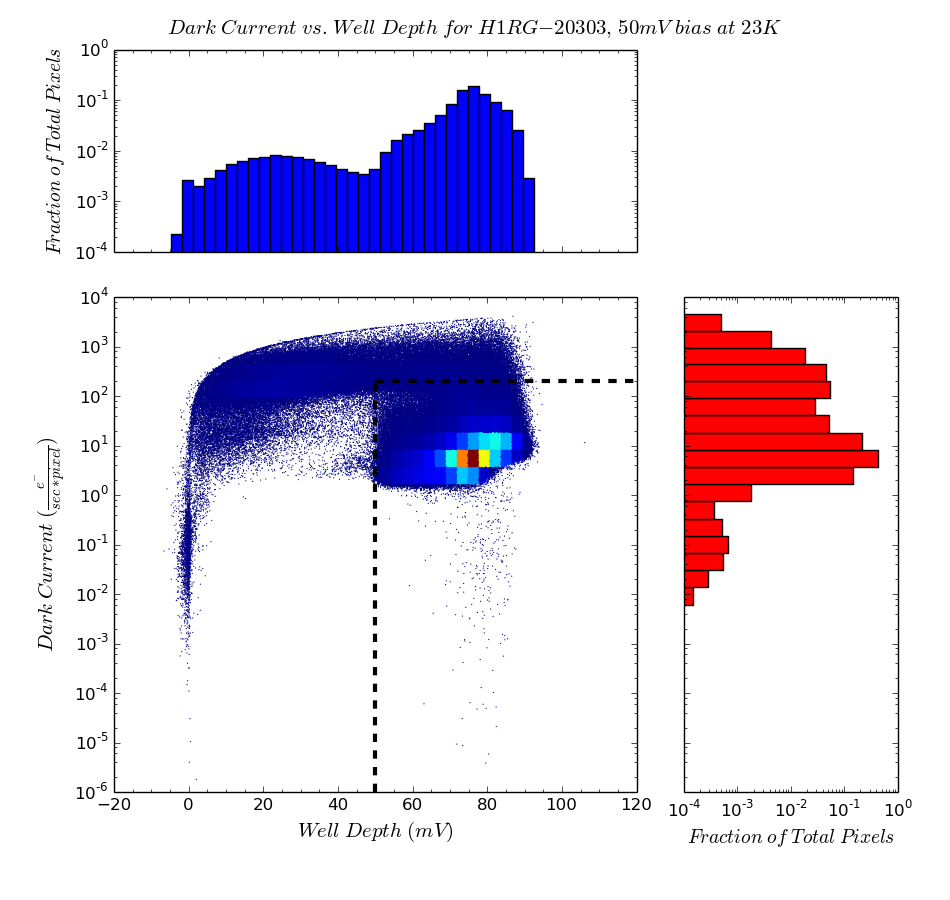}
\end{tabular}
\end{center}
\caption 
{ \label{fig:Ivswell_50mV_23K}
Current in the dark \textit{vs.} well depth distribution per pixel for H1RG-20303 at a temperature of 23 K and an applied bias of 50 mV. The vertical dashed line corresponds to well depth of $\sim$14 $ke^-$. Note that the well depth corresponds to the actual initial detector bias.}
\end{figure}

Table \ref{tab:median dark current and well depth} shows the median dark current and well depth for H1RG-20303 at all tested temperatures (23-30 K) and applied biases (50-350 mV). Figure \ref{fig:Ivswell_50mV_23K} shows the dark current \textit{vs.} well depth distribution per pixel at a temperature of 23 K and an applied bias of 50 mV. This distribution is used to identify pixels that have large initial dark current and/or low well depth as inoperable. The pixels below and to the right of the dashed lines are considered to be operable pixels since they have relatively low dark currents (below our fiducial requirement of 200 $e^-/s$ based on the NEOCam requirements for their LW10 arrays) and sufficient well depth (arbitrarily chosen to include the majority of pixels). Low well depth and low dark currents are indicators that those pixels have very large dark currents and have debiased appreciably between the reset and the first time the pixel is addressed.

\begin{figure}
\begin{center}
\begin{tabular}{c}
\includegraphics[scale=0.17]{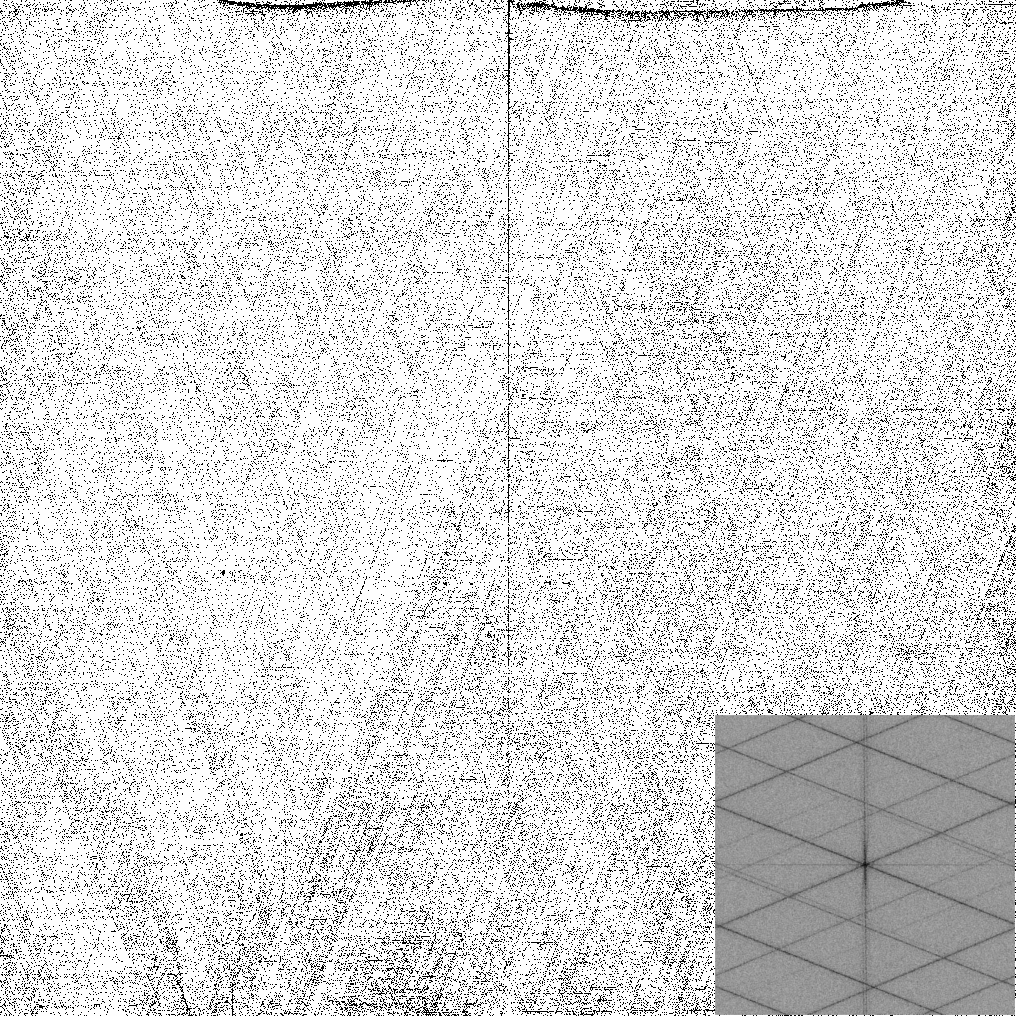}
\end{tabular}
\end{center}
\caption 
{ \label{fig:oper_50mV_23K}
Operability map for H1RG-20303 at a temperature of 23 K and applied bias of 50 mV, where inoperable pixels are shown in black. Operable pixels (87.3\%) have dark currents below 200 $e^-/s$ and well depths greater than 14 $ke^-$.} 
\end{figure}

Figure \ref{fig:oper_50mV_23K} shows the operability map of the array at 23 K and 50 mV of applied bias, where the black pixels (inoperable) denote those that have dark currents above 200 $e^-/s$, well depths below $\sim$14 $ke^-$, or both. A collection of inoperable pixels form a cross-hatching pattern in Fig. \ref{fig:oper_50mV_23K} lying along directions parallel to the set of three cross-hatching lines that are formed by the lattice mismatch between HgCdTe and the CdZnTe substrate\cite{Martinka2001,Chang2008}. This cross-hatching pattern was also observed in all four LW13 arrays.

The FFT of the operability map (lower right corner in Fig. \ref{fig:oper_50mV_23K}) shows the cross-hatching pattern rotated by \ang{90} as a set of parallel lines in three distinct directions. The prominent vertical line in the FFT corresponds to the horizontal cross-hatching pattern seen in this operability image. The large number of bad pixels at the top of the array indicates an issue with the HgCdTe growth in that region. The vertical line of inoperable pixels in the operability map may be due to an issue with the multiplexer: it corresponds to the faint horizontal line in the FFT, and is not relevant to our detector evaluation. The Hough transform was applied to the FFT of the operability map to estimate the angles of the cross-hatching pattern, and are consistent with those found by other authors\cite{Martinka2001,Chang2008}. The angle between the lines parallel to the two diagonal cross-hatching pattern is \ang{44.5}, corresponding to the $\left[\overline{2}31\right]$ and $\left[\overline{2}13\right]$ directions in the crystal lattice. The third cross-hatching pattern is parallel to the $\left[01\overline{1}\right]$ direction, and is rotated clock-wise from the horizontal axis by \ang{0.5}.

\section{Dark Current Model}
\label{sect:dark current results}

\begin{figure}
\begin{center}
\begin{tabular}{c}
\includegraphics[scale=0.45]{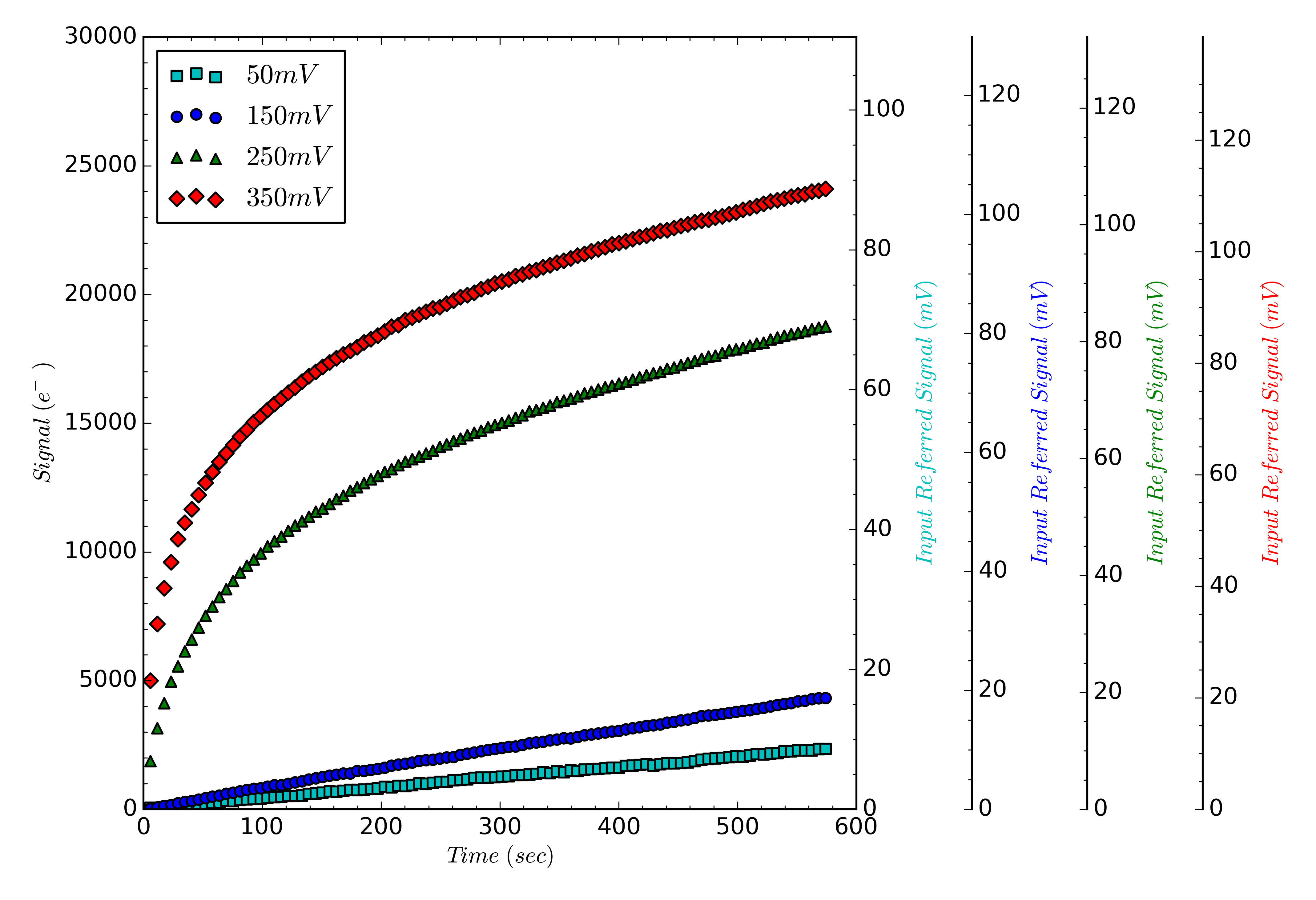}
\end{tabular}
\end{center}
\caption 
{ \label{fig:sutr}
Discharge history for pixel [527, 459] from H1RG-20303 at a temperature of 23K and at applied reverse bias of 50, 150, 250, and 350 mV in the dark. On the left y-axis scale we show the signal in electrons, while the right y-scale axis shows the input referred signal in mV. The four different scales on the right correspond to the different diode capacitance measured (from left to right) at 50, 150, 250, and 350 mV of applied biases. The initial dark currents (and actual initial bias) for the four SUTR curves at 50, 150, 250, and 350 mV are: 4 $e^-/s$ (77 mV), 8 $e^-/s$ (181 mV), 322 $e^-/s$ (282 mV), and 861 $e^-/s$ (312 mV) respectively.} 
\end{figure} 

Figure \ref{fig:sutr} shows the discharge history for one pixel at a temperature of 23 K and all four applied biases in the dark. As mentioned in the Sect. \ref{sect:theory}, quantum tunneling dark currents are a strong function of bias, where this effect can be seen in the large slope increase at the beginning of the dark signal \textit{vs.} time curve when the applied bias is increased from 150 to 250 mV and continues to increase at higher biases. To determine the mechanisms of dark current that are present/dominate as the diodes debias, the models presented in sect. \ref{sect:theory} are fitted to dark current data as functions of temperature and bias.

The first step in fitting the dark current mechanism models is to estimate the band-to-band tunneling. Both tunneling currents vary exponentially with the band gap energy and the electric field in the junction, $E_g^{3/2}/E \equiv \beta$, where this is the only parameter to fit for band-to-band tunneling. Once band-to-band tunneling has been estimated, diffusion and G-R currents are fitted to the higher temperature dark current data, which is expected to be dominated by these thermally generated dark currents. The minority carrier lifetimes for diffusion and G-R are optimized to fit both of these dark current components simultaneously ($\tau_h$ and $\tau_{GR}$ respectively), in addition to the trap energy ($E_{t_{gr}}$) that contributes the most to G-R. If the fitted band-to-band tunneling, diffusion, and G-R do not compensate for all of the dark current behavior over the entire I-T and I-V curves, then a light leak and trap-to-band tunneling current are fitted. The light leak is fitted as a constant current, while the trap energy and density ($E_t$ and $n_t$ respectively) are optimized to fit trap-to-band tunneling.

H1RG-20303 appears to have a trap-to-band tunneling contribution at the largest applied bias of 350 mV, where only 3-5 data points at the largest bias would follow a band-to-band tunneling current trend exclusively. With an applied bias of 350 mV, band-to-band tunneling current is large enough to debias the array to a median of 306 mV between the reset and the first time the array is read out. In order to get an accurate estimate of $\beta$, dark current and well depth data were taken with 400 mV of applied bias at 23 K with an integration time of 167 ms (to reduce the time between reset and the first frame, thereby reducing the amount the array debiases), addressing only the 32 rows that were used to take warm-up data. This 400 mV dark SUTR data provided several more points that followed the trend of band-to-band tunneling.

\begin{figure}
\begin{center}
\begin{tabular}{c}
\includegraphics[scale=0.39]{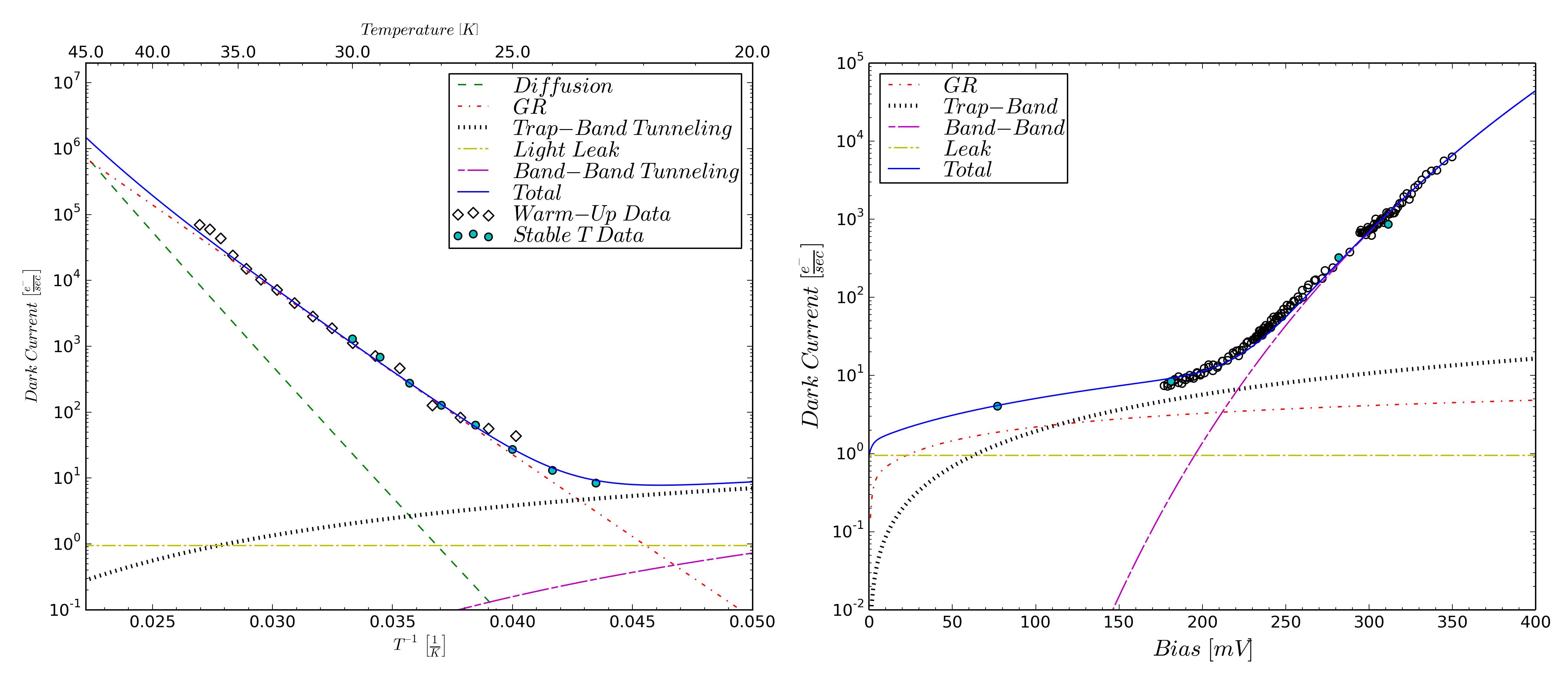}
\\
(a) \hspace{7.4cm} (b)
\end{tabular}
\end{center}
\caption 
{ \label{fig:Imodel}
(a) Dark current \textit{vs.} temperature with an applied bias of 150 mV for pixel [527, 459] in H1RG-20303. (b) Dark current \textit{vs.} bias (at 23 K). The solid cyan circle data points are the initial dark current and initial detector bias for dark SUTR curves taken with applied biases of 50, 150, 250, and 350 mV. 400 mV dark current and well depth data was taken for only 32 rows of pixels with an integration time of 167 ms, allowing us to measure dark currents at higher biases before large dark currents debiased the pixels. The 167 ms data corresponds to the open circle data to the right of the cyan data point at 311 mV.} 
\end{figure}

Figure \ref{fig:Imodel} shows the analysis of the dark current at different biases and temperatures for the pixel shown in Fig. \ref{fig:sutr}. At biases greater than about 250 mV, the dark current for the pixel shown in Fig. \ref{fig:Imodel} is dominated by band-to-band tunneling, while at lower biases and low temperatures a combination of G-R, defect assisted tunneling currents, and a small light leak are the dominant dark current components. At the higher temperatures (up to 37 K), dark current from G-R appears to dominate. The optimized light leak and dark current parameters to fit the data are shown in table \ref{tab:parameters}.

\begin{figure}
\begin{center}
\begin{tabular}{c}
\includegraphics[width = 0.97\textwidth]{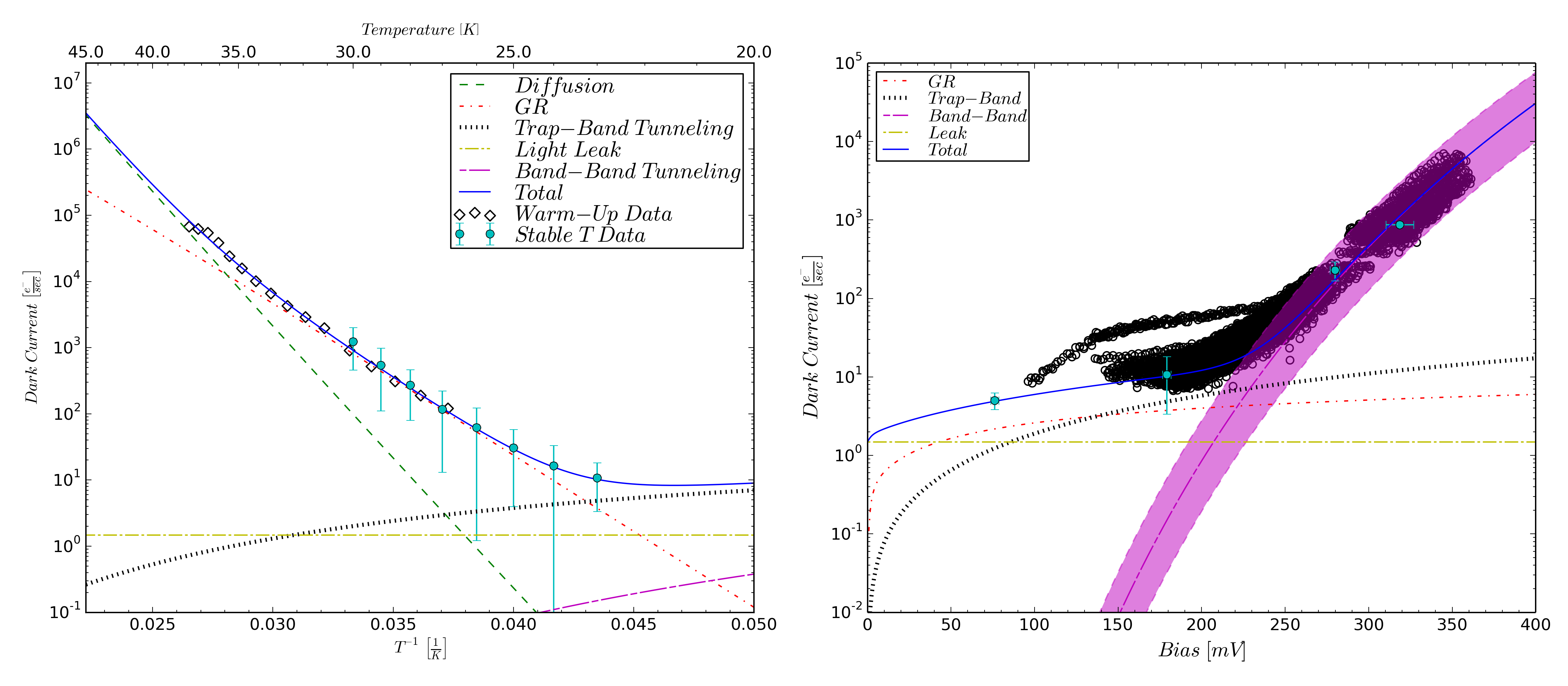}
\\
\hspace{.5cm} (a) \hspace{7.cm} (b)
\end{tabular}
\end{center}
\caption 
{ \label{fig:20303_I_model_36pix}
(a) Median dark current \textit{vs.} temperature data for 36 operable pixels in H1RG-20303. The median detector bias of 179 mV, corresponding to the 23 K stable data point, was used to fit the dark current models.The error bars on the median stable temperature data correspond to $\pm$ one standard deviation of the mean in the dark current values for the pixels. (b) Individual dark current \textit{vs.} bias curves for the same 36 pixels obtained from the 250, 350, and 400 mV SUTR curves taken to measure dark current and well depth for this array at a temperature of 23 K. In addition to the individual I-V curves, the median initial dark current and well depth of SUTR curves at 150, 250, and 350 mV were added, where the error bars correspond to $\pm$ one standard deviation of the mean in the measurements for the pixels. The shaded region in (b) corresponds to the band-to-band tunneling calculated from the $\beta$ value $\pm$ two standard deviations from the mean, while the single band-to-band tunneling curve corresponds to that of the mean $\beta$ value of the 36 pixels.}
\end{figure}

Similar to the LW13 devices, to show the uniformity of band-to-band tunneling in H1RG-20303, the parameter $\beta$ was fitted to operable pixels (to avoid larger trap-to-band tunneling contributions at larger biases) in the central 32$\times$400 pixel region of the array (10,700 pixels).
Figure \ref{fig:20303_I_model_36pix} shows dark current data for 36 randomly selected operable pixels from the region where $\beta$ was estimated. The boundary of the shaded region in the I-V curve corresponds to the band-to-band tunneling calculated from the $\beta$ value $\pm$ two standard deviations from the mean of the fitted $\beta$ parameter distribution. The cyan data points correspond to the median initial dark current of the 36 pixels, where the error bars correspond to one standard deviation from the mean, while the warm-up data is also the median of the 36 pixels. The dark current model was fitted to the median values on both the I-T and I-V curves, where the same general behavior was observed as in the fit to the individual pixel in Fig. \ref{fig:Imodel}. The individual I-V curves for the 36 pixels are added to show the uniformity of band-to-band tunneling. The same general behavior was observed among many of the individual operable pixels that were analyzed, where at the larger biases and temperature of 23 K, band-to-band tunneling is the main dark current mechanism present. At lower biases band-to-band tunneling decreases significantly and the current in the dark is affected by a combination of G-R, trap-to-band tunneling, and the leak in the test dewar. At the higher temperatures of the available warm-up data, G-R is the main source of dark current. Higher temperature warm-up data is needed to better characterize diffusion current.

\begin{table}
\centering
\caption{Optimized dark current parameters to model the dark current behavior shown in Figures \ref{fig:Imodel} and \ref{fig:20303_I_model_36pix}.}
\label{tab:parameters}
\begin{tabular}{c|c|c|}
\cline{2-3}
\multicolumn{1}{c|}{} & Pixel [527, 459] & Median of 36 Pixels \\ \hline
\multicolumn{1}{|c|}{$\beta\ \left[\frac{eV^{3/2}}{V/cm}\right]$} & $5.38\times10^6$ & $5.44\times10^6$ \\ \hline
\multicolumn{1}{|c|}{$\tau_h\ \left[sec\right]$} & $8.1\times10^{-7}$ & $2\times10^{-7}$ \\ \hline
\multicolumn{1}{|c|}{$\tau_{GR}\ \left[sec\right]$} & $1.1\times10^{-5}$ & $4.7\times10^{-5}$ \\ \hline
\multicolumn{1}{|c|}{$E_{t_{gr}}\ \left[eV\right]$} & $4.6\times10^{-2}$ & $4.3\times10^{-2}$ \\ \hline
\multicolumn{1}{|c|}{$n_t\ \left[m^{-3}\right]$} & $3\times10^6$ & $3.7\times10^6$ \\ \hline
\multicolumn{1}{|c|}{$E_t\ \left[eV\right]$} & $6.9\times10^{-2}$ & $6.8\times10^{-2}$ \\ \hline
\multicolumn{1}{|c|}{Light Leak $\left[e^-/sec\right]$} & 0.95 & 1.5 \\ \hline
\end{tabular}
\end{table}

It is also important to note that the light leak and trap-to-band tunneling that were fitted here may not have a unique solution. The initial guess given to the least squares optimizing function will result in different fitted parameters that fit the data well. Regardless of the fitted parameters for these two components, there is a dark current component that is bias dependent, which can be explained by trap-to-band in both cases (the single pixel, and the fit to the median of 36 pixels). The light leak would not vary with bias to compensate for this bias-dependent behavior. While G-R does have a small bias dependence, it is not enough to compensate for the observed behavior.

There appears to be an increase in trap-to-band tunneling among well behaved pixels in this array over the LW13 devices. All of the individual pixels that were studied showed a similar trap-to-band tunneling dependence as the majority of the pixels shown here, but a more detailed study of the dark current is needed to determine if this is an array-wide increase in trap-to-band tunneling. Unlike band-to-band tunneling, not only are there more parameters to fit for trap-to-band tunneling, but the parameters for this mechanism must be optimized simultaneously with the light leak (may not be uniform across the array), which as was mentioned above can yield a non-unique solution and in some cases the optimization of the parameters will not converge. Additionally, though the uniform trap density model of trap-to-band tunneling (no modification to eqn. \ref{eq:trap-to-band}) is sufficient to fit the behavior of many of the individual plotted pixels, a few of the pixels (one shown in Fig. \ref{fig:20303_I_model_36pix}) do appear to show the soft breakdown that precedes the onset of trap-to-band tunneling, and was seen in inoperable pixels in the LW13 devices and the LW10 devices studied in Wu (1997) and Bacon (2006). To model the soft breakdown, the addition of the parametric fit corresponding to a changing trap density (eqn. \ref{eq:n_t}) would increase the total number of parameters to fit for trap-to-band tunneling to five, making the optimization of the parameters more difficult.

Even with this apparent increase in trap-to-band tunneling, the results obtained for this array are very encouraging as it shows an improvement in the array design by TIS in increasing the $\beta$ parameter (see Cabrera et al. (2019)\cite{Cabrera19}), therefore reducing tunneling currents. The extrapolated band-to-band tunneling current fitted to H1RG-18509 (LW13 array designed to reduce tunneling currents) to a device with the same cutoff wavelength as H1RG-20303 would be $\sim 5\times10^5 e^-/sec$ with an actual reverse bias of 350 mV at a temperature of 23 K, about two orders of magnitude larger than what was measured for the pixel in Fig. \ref{fig:Imodel}.

\section{Summary}
\label{sect:summary}

Preliminary dark current analysis of H1RG-20303 shows the successful improvement in array design by TIS to further reduce tunneling currents, driven by the results from the LW13 phase arrays. Up to a temperature of 25 K and an applied reverse bias of 50 mV, $\sim$ 84.4\% of the pixels have dark currents less than 200 $e^-/s$ and well depths of at least 14 $ke^-$ (77.6\% at 26 K), where the percentage is expected to increase for arrays with the target wavelength of 15 $\mu m$.

Future work will involve a more detailed I-T and I-V analysis of an ensemble of pixels to better assess the uniformity or lack thereof in the different dark current mechanisms for this array. Two more LW15 arrays will also be characterized.

\acknowledgments
The University of Rochester group acknowledge support by NASA grant NNX14AD32G S07. M. Cabrera acknowledges the NASA grant, New York Space grant, and the Graduate Assistance in Areas of National Need (GAANN) grant for partially supporting his graduate work.

\bibliography{refs}
\bibliographystyle{spiebib}

\end{document}